\documentclass[aps,prl,twocolumn,groupedaddress,showpacs]{revtex4}
\usepackage{bm}
\usepackage{dcolumn}
\usepackage{graphicx}
\usepackage{subfigure}
\newlength{\myw}

\begin{document}

\title{Nanotubular Boron-Carbon Heterojunctions}
\author{Jens Kunstmann}
\email{kunstmann@physik.uni-greifswald.de}
\author{Alexander Quandt}
\affiliation{Institiut f\"ur Physik, Domstra\ss e 10a,
Ernst-Moritz-Arndt-Universit\"at Greifswald, 17489 Greifswald,
Germany}

\date{\today}

\begin{abstract}
Linear nanotubular boron-carbon heterojunctions are systematically
constructed and studied with the help of \textit{ab initio} total
energy calculations. The structural compatibility of the two classes 
of materials is shown, and a simple recipe that determines 
all types of stable linear junctions is illustrated in some detail. 
Our results also suggest the compatibility of various technologically 
interesting types of nanotubular materials, leading to novel types 
of nanotubular compound materials, and pointing out the possibility 
of wiring nanotubular devices within heterogeneous nanotubular 
networks.
\end{abstract}

%
%
%
%
\pacs{81.07.Lk, 81.07.De, 73.22.-f, 61.46.+w}

\maketitle


\section{Introduction}
Carbon nanotubes (CNTs) \cite{iijima:91} have a number of fascinating
properties, which could make them the material of choice for the 
future miniaturization of electronic devices: They are good thermal 
conductors, exhibit a good thermal resilience, are ductile and 
superb field emitters \cite{collins:00}. Furthermore their electronic 
properties range from metallic to semi-conducting, depending on their 
chiralities and radii \cite{dresselhaus:SFCN}. 

Besides there are many other nanotubes made from inorganic 
materials \cite{tremel_1999_acheint,tenne_2001_cnt}, like 
BN \cite{chopra_1995_sci} or MoS$_2$ \cite{tenne_1992_nat} 
nanotubes which have already been synthesized, 
or materials that have been predicted by theory  
like metal-boron nanotubes \cite{quandt:01,zhang_2002_prl} 
or the structurally related CaSi$_2$ nanotubes \cite{gemming_2003_prb}.
Another promising new type of nanotubular materials are nanotubes 
consisting of pure boron (B-nanotubes) \cite{boustani:97,gindulyte_1998_ic}.
The stability and the mechanical properties of B-nanotubes (BNTs) 
would be quite similar to C- and BN-nanotubes, but all BNTs should 
be metallic, independent of their chirality \cite{boustani:99}. 
Very recently Ciuparu \textit{et al.} \cite{ciuparu_2004_jpcb} 
successfully synthesized these nanotubes and thus confirmed 
the existence of BNTs, after similar efforts had already lead 
to the discovery of novel types of boron nanowires by various 
other groups \cite{cao:01,wu_2001_admat,otten:02,zhang_2002_chemcom}.


With such a large number of nanotubular systems available (a number  
which is very likely to further increase in the near future), possible 
nano\-technologies are most likely to be based on this \textit{variety} 
of nanotubular materials rather than just CNTs (see \cite{chico:96,yao:99}). 
The functionalities of such \textit{heterogeneous nanotubular networks} 
would mainly arise from their composite character.

But before one can seriously start any discussion about the future 
impact of heterogeneous nanotubular networks, it will be necessary 
to explore the stability of their key elements, first: 
the interfaces between different nanotubular materials. 
In the following, we will, for the first time, study the 
compatibility of different nanotubular materials in a 
\textit{systematic fashion} by examining the structural and electronic 
properties of suitable model junctions between those systems. 

To this end we decided to focus our studies on boron--carbon 
nano\-junctions, which may be seen as a structural paradigm for a 
large class of junctions, comprising most of the non--carbon based 
nanotubular materials mentioned above. On the basis of three exemplary 
structures, we will demonstrate the compatibility of CNTs and BNTs. 
Due to the exemplary character of the model junctions chosen for 
this study, our results also suggest the existence of similar junctions 
between a large number of related nanotubular materials.

\begin{figure*}
\centering
\subfigure(a){\includegraphics[height=10cm]{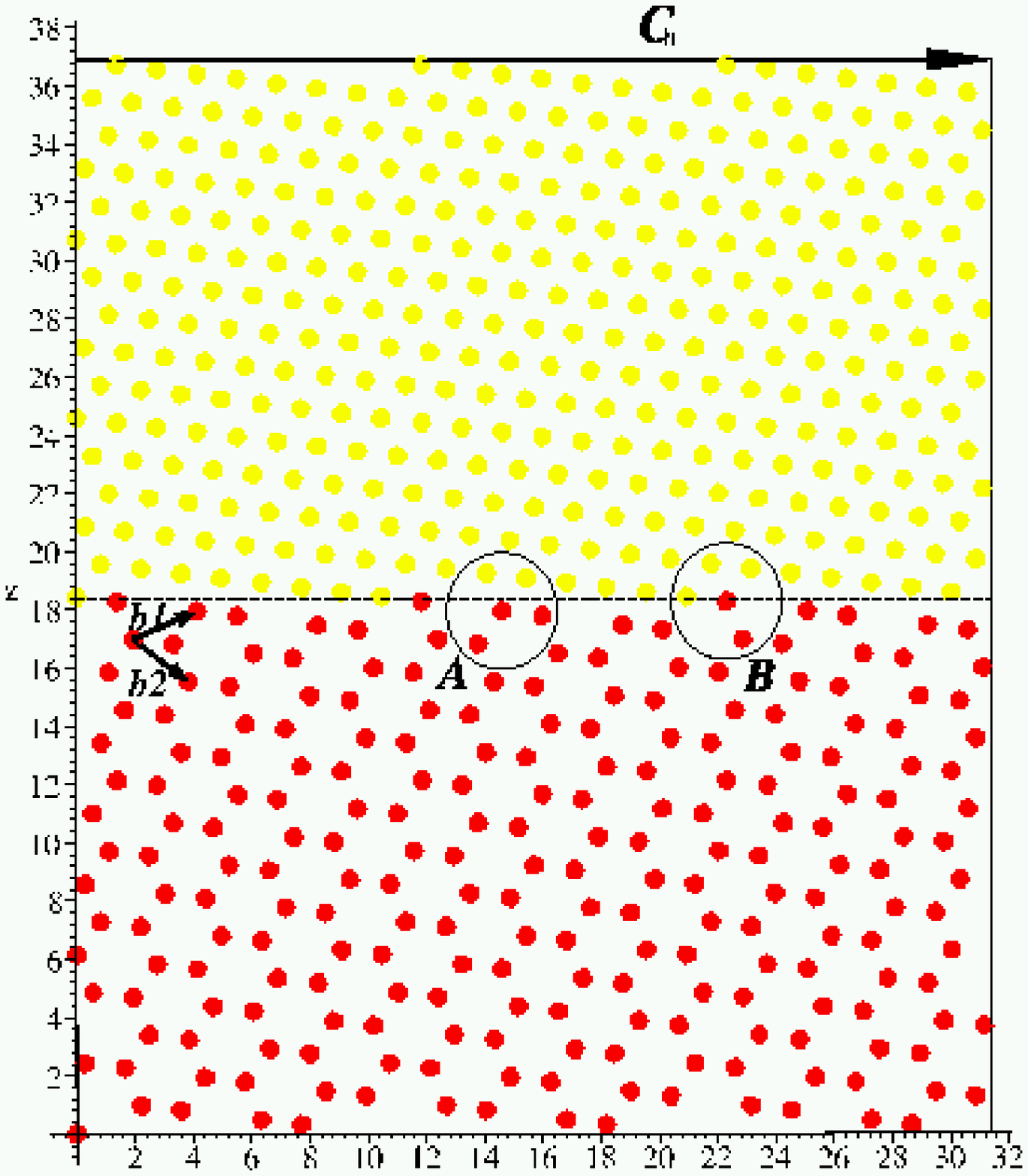}}
\subfigure(b){\includegraphics[height=10cm]{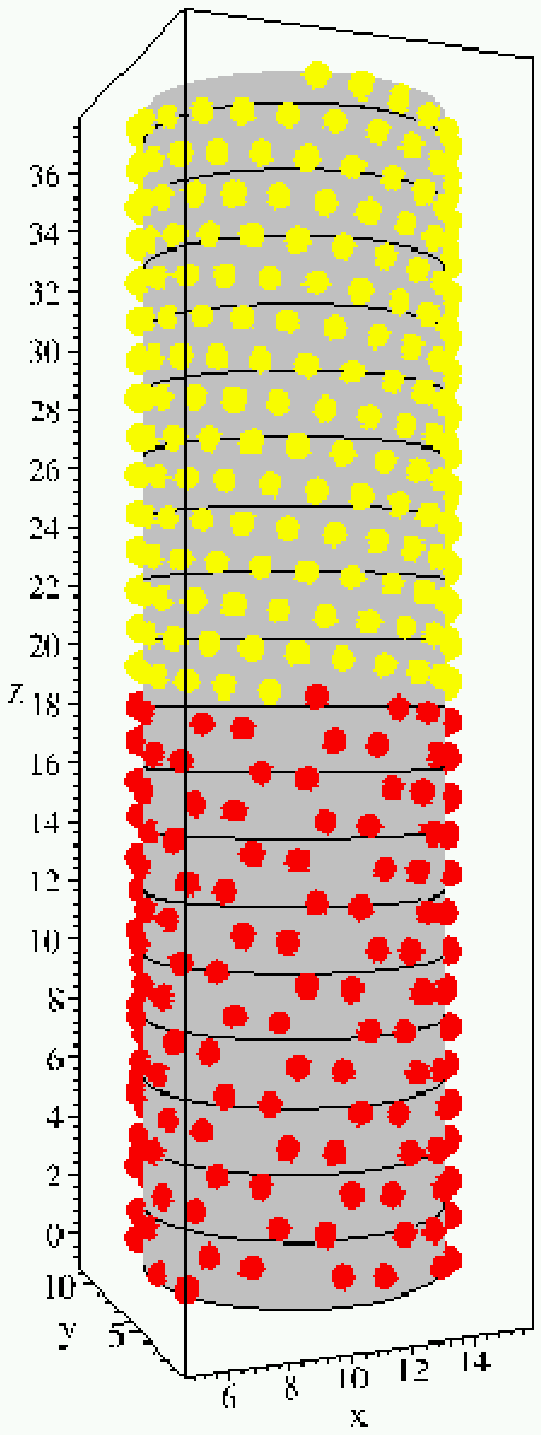}}
\caption{\label{fig:illu}
Basic construction of B-C junctions.
(a): (9,6) sheet, units are \AA, light atoms are boron, dark atoms are carbon. 
Also indicated are the chiral vector $\bm{C}_{\mathrm h}$ and the basis 
vectors of a honeycomb lattice $\bm{b_1}$,$\bm{b_2}$. The encircled parts
emphasize the two structure elements $A$ and $B$ that appear in the 
transition region, and the dashed line separates the boron part from the 
carbon part. (b): chiral nanojunction constructed from (a) by rolling up 
the sheet and gluing it at the vertical sides, where the grey cylinder 
should emphasize the resulting tubular structure.}
\end{figure*}


\section{Nanotubular Heterojunctions}
Nanotubes are geometrically constructed by rolling up a
rectangular sheet that has been cut from a planar structure. For
CNTs this planar structure will be the honeycomb lattice
\cite{dresselhaus:SFCN}, while for BNTs the planar
reference structure will be a (puckered) hexagonal lattice
\cite{boustani:99}.

In order to approach the rather complex problem of heterogeneous nanotubular 
junctions we restricted our examinations to linear B-C junctions with 
constant chirality, where both parts of the junctions have similar radii. 
The sought B-C nanojunctions (and all other structures considered in this paper) 
are directly related to the structure of pure CNTs, as they can obviously 
be constructed from a single sheet. Therefore we may also classify them  
in a standard fashion employing a pair of integers (N,M) which specify 
the chiral vector $ \bm{C}_{\mathrm h} = N \bm{b_1} + M \bm{b_2} $. 
Here $\bm{b_1}$ and $\bm{b_2}$ are basis vectors for a honeycomb 
lattice (see Fig.\ref{fig:illu}). 

Now the general prescription to create our sample B-C junctions
goes as follows: A rectangular sheet is cut from a honeycomb
lattice, such that its horizontal sides equal the chiral vector,
and the vertical sides (dubbed $z$-direction in the following) 
being even multiples of a basic period, depending on the chirality 
of the tube \cite{dresselhaus:SFCN}. An imaginary horizontal line 
in the middle of the sheet separates the upper and the lower part. 
The lower part consists of the original honeycomb lattice and is 
decorated with carbon atoms at each lattice site. The upper part 
is transformed into a hexagonal lattice after inserting another 
lattice point at the center of the honeycombs. Next, this structure 
is decorated with boron atoms. Finally the sheet is rolled up and glued 
along its vertical sides (see Fig. \ref{fig:illu}). 

For the stability of these junctions it will be crucial to understand 
the local atomic configurations close to the interface. Using the ge\-neration 
scheme described above, only two different structure elements, dubbed 
$A$ and $B$ in the following, may form close to the transition region 
(emphasized in Fig. \ref{fig:illu}). Thus a detailed study of the basic 
structural and chemical properties related to the formation of $A$ and $B$ 
will be sufficient to understand the stability of any linear and chiral 
nanojunction. These structure elements located around the interface regions 
will imply a planar coordination number of four for the constituent carbon 
atoms. Such coordinations are not found in pure CNTs, but higher planar 
coordinations of carbon atoms in contact with boron atoms have already 
been reported by Exner \textit{et al.} \cite{exner:00} for small BC-clusters.
We also note that non-chiral armchair and zigzag systems contain only 
one type of structure element $A$ or $B$, and in order to be able 
to study their effects in a systematic fashion, we will restrict 
ourselves to armchair and zigzag types of nanojunctions.


\section{Theoretical Methods}
 
Due to its electron deficient character \cite{pauling_1960_ncb} boron has a complicated and versatile chemistry. The only theoretical tools that allow to describe its chemistry properly are first principles calculations \cite{boustani_1997_prb}.

\begin{figure}
\centering
  \includegraphics[width=0.6\linewidth]{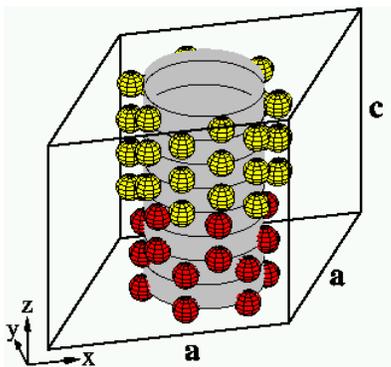}        
\caption{\label{fig:UnitCell}
The unit cell for hexagonal tubular bundles (ropes), with its lattice constants $a$ and $c$, and the atomic decoration for the unrelaxed approximant Z1 (see text). Light atoms are boron, dark atoms are carbon and the grey cylinder again emphasizes the tubular structure of the junction.
}
\end{figure}

In order to carry out \textit{ab initio} type of simulations for nanotubular 
compound systems, we have to construct a solid composed of suitable unit cells 
containing those junctions. Within the xy-plane, we arrange the tubular 
junctions side by side on a hexagonal lattice of lattice constant $a$. 
Therefore we will simulate bundles (ropes) of linear junctions rather than 
single free-standing nanojunctions. In the z-direction we simply pile up our unit cells (with lattice 
constant $c$). The whole procedure will lead to tubular structures that are 
a linear array of B-C and C-B transitions (see Fig. \ref{fig:UnitCell}).
Due to the periodic boundary conditions these systems can at best 
\textit{approximate} single nanojunctions that would be present in 
the limit $a,c \rightarrow \infty$. Nevertheless the small 
approximants considered in this paper are sufficient to understand 
the properties of the structure elements $A$ and $B$, as shown below. 

The numerical calculations themselves were carried out using the VASP
\textit{ab initio} package, version 4.4.5 \cite{kresse:96-1,kresse:96-2}. 
The latter is a density functional theory (DFT) \cite{kohn:65} based 
{\em ab initio} code using plane wave basis sets and a (super)cell 
approach to model solid materials. During all simulations, the 
electronic correlations were treated within the local-density 
approximation (LDA \cite{ceperley:80}), and the ionic cores of the 
system were represented by ultrasoft pseudopotentials \cite{vanderbilt:90}. 
With the help of the VASP program, one can determine interatomic forces and 
relax all degrees of freedom for a given decorated unit cell in order to
detect atomic configurations which correspond to (local) minima on
the total energy hypersurfaces. In order to perform these kinds of structure 
optimizations we employed a preconditioned conjugate gradient algorithm 
\cite{teter:89} and let \textit{all} degrees of freedom relax, i.e. the 
complete set of atomic configurations as well as the unit cell parameters. 
The total energy as well as the $k$-point sampling were converged such that 
changes in the total energies were less than $10^{-3}$ eV and interatomic 
forces were less than 0.05 eV/\AA.

\begin{table*}
\caption{\label{tab:data} Structure data and energies for nanojunctions and 
reference structures. (N,M): nanotube structure type, C: number of carbon 
atoms per unit cell, B: number of boron atoms per unit cell, (a,c): lattice 
constants $a$ and $c$ of unit cell in \AA, $a_{\mathrm{C-C}}$, $a_{\mathrm{B-B}}$, 
$a_{\mathrm{B-C}}$: range of bond lengths in \AA \ of C-C, B-B, and B-C bonds, 
respectively, $E_{\mathrm{coh}}$: cohesive energies in eV/atom, 
$E_{\mathrm{coh}}^{\mathrm{ref}}$: reference cohesive energy in eV/atom 
(see text).} 
 \newcolumntype{e}[0]{D{,}{,}{5.5}}
 \newcolumntype{f}[0]{D{,}{,}{2.2}} 
 \begin{ruledtabular}
 \begin{tabular}{lfccecccddd}
 System      & (N,M) & C & B &(a,c)& $a_{\mathrm{C-C}}$ & $a_{\mathrm{B-B}}$ & $a_{\mathrm{B-C}}$ & \multicolumn{1}{c}{$E_{\mathrm{coh}}$} & \multicolumn{1}{c}{$E_{\mathrm{coh}}^{\mathrm{ref}}$}
  & \multicolumn{1}{c}{$E_{\mathrm{coh}}^{\mathrm{ref}} - E_{\mathrm{coh}}$}\\
 \hline
 Armchair 1 (A1) & (6,6) & 24 & 36 & (12.06, 5.54)  & $1.40 \dots 1.45$ & $1.65 \dots 1.85$ & $1.53 \dots 1.57$ & 8.079 & 8.212 & 0.13\\
 Zigzag 1 (Z1)   & (6,0) & 24 & 36 &  (8.32, 8.95)  & $1.40 \dots 1.45$ & $1.61 \dots 1.96$ & $1.48 \dots 1.74$ & 7.964 & 8.038 & 0.07\\
 Armchair 2 (A2) & (6,6) & 36 & 84 & (11.93, 10.70) & $1.40 \dots 1.45$ & $1.58 \dots 1.95$ & $1.49 \dots 1.72$ & 7.839 & 7.910 & 0.07\\
 \hline
 C-Nanotube & (6,6) & 24 & $-$  &(11.17,2.45) & 1.41 & $-$ & $-$ & 10.023 & -&-\\
 B-Nanotube & (6,6)\footnote{relaxed structure from \cite{quandt:01}} & $-$  & 36 &(12.42,2.84)& $-$ & $1.58 \dots 1.86$ & $-$ & 7.004 & -&- \\
 C-Nanotube & (6,0) & 24 & $-$  &(8.20,4.22)& $1.39 \dots 1.43$ & $-$ & $-$ & 9.791 & -&-\\
 B-Nanotube & (6,0) & $-$  & 12 &(8.18,1.65)& $-$ & $1.65 \dots 1.78$ & $-$ & 6.870 & -&-\\

 \end{tabular}
 \end{ruledtabular}
\end{table*}

It turned out that the energy hypersurfaces of our structures were 
rather complex and full of local minima. In order to approach the 
most stable structures, we developed a procedure where the starting
configurations were prerelaxed with lower numerical precision
until a (local) minimum was found, and afterwards we continued
the relaxations with optimal precision. A reduced precision leads to
somewhat imprecise interatomic forces, but we found that such a
procedure would result in the scanning of the energy hypersurface
over a wider range. 

\begin{figure}
\setlength{\myw}{.49\linewidth}
\parbox{\myw}{(a) Armchair (6,6) CNT}
\parbox{\myw}{(b) Armchair (6,6) BNT}
\subfigure{\includegraphics[angle=-90,width=\myw]{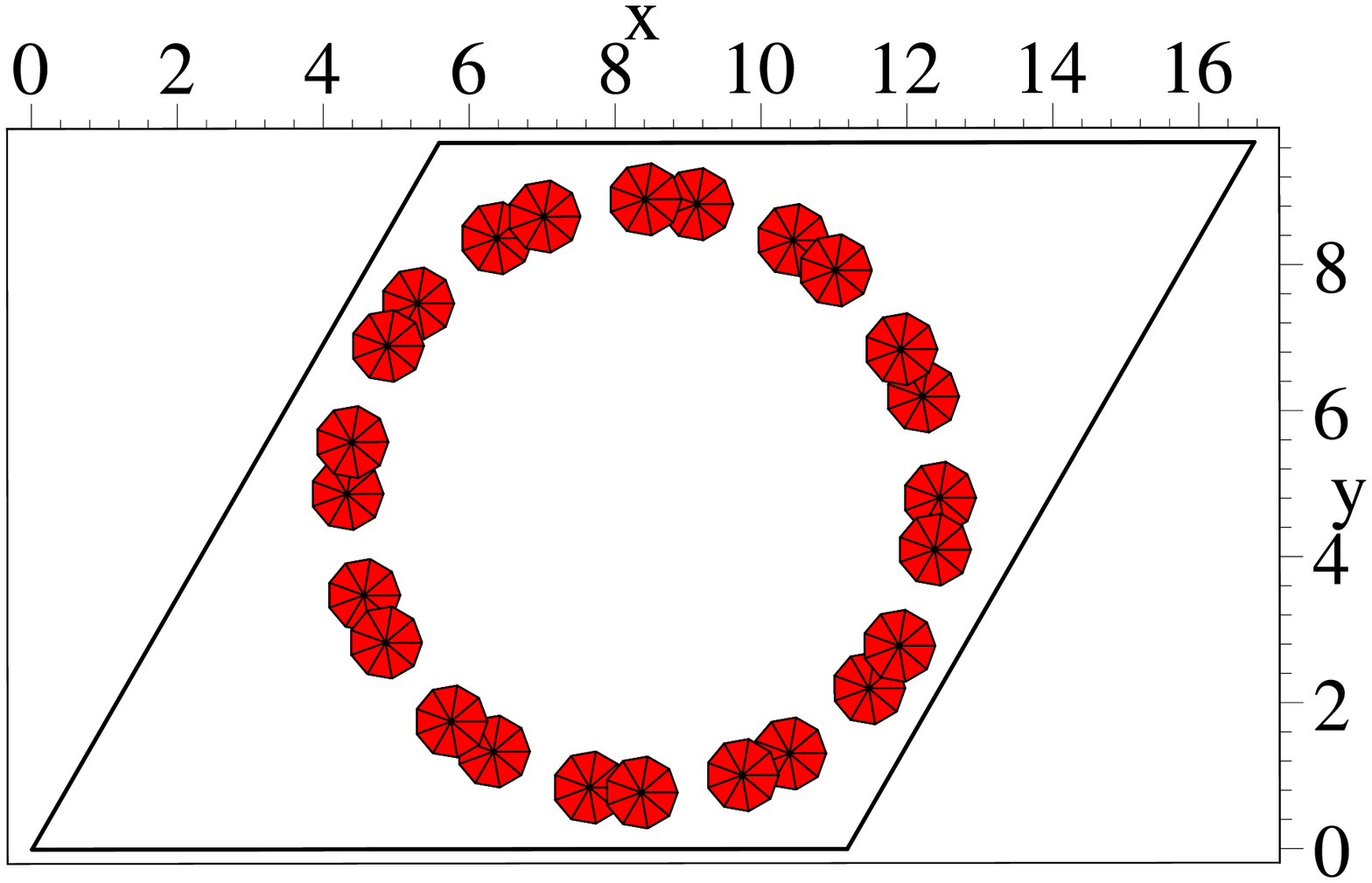}}
\subfigure{\includegraphics[angle=-90,width=\myw]{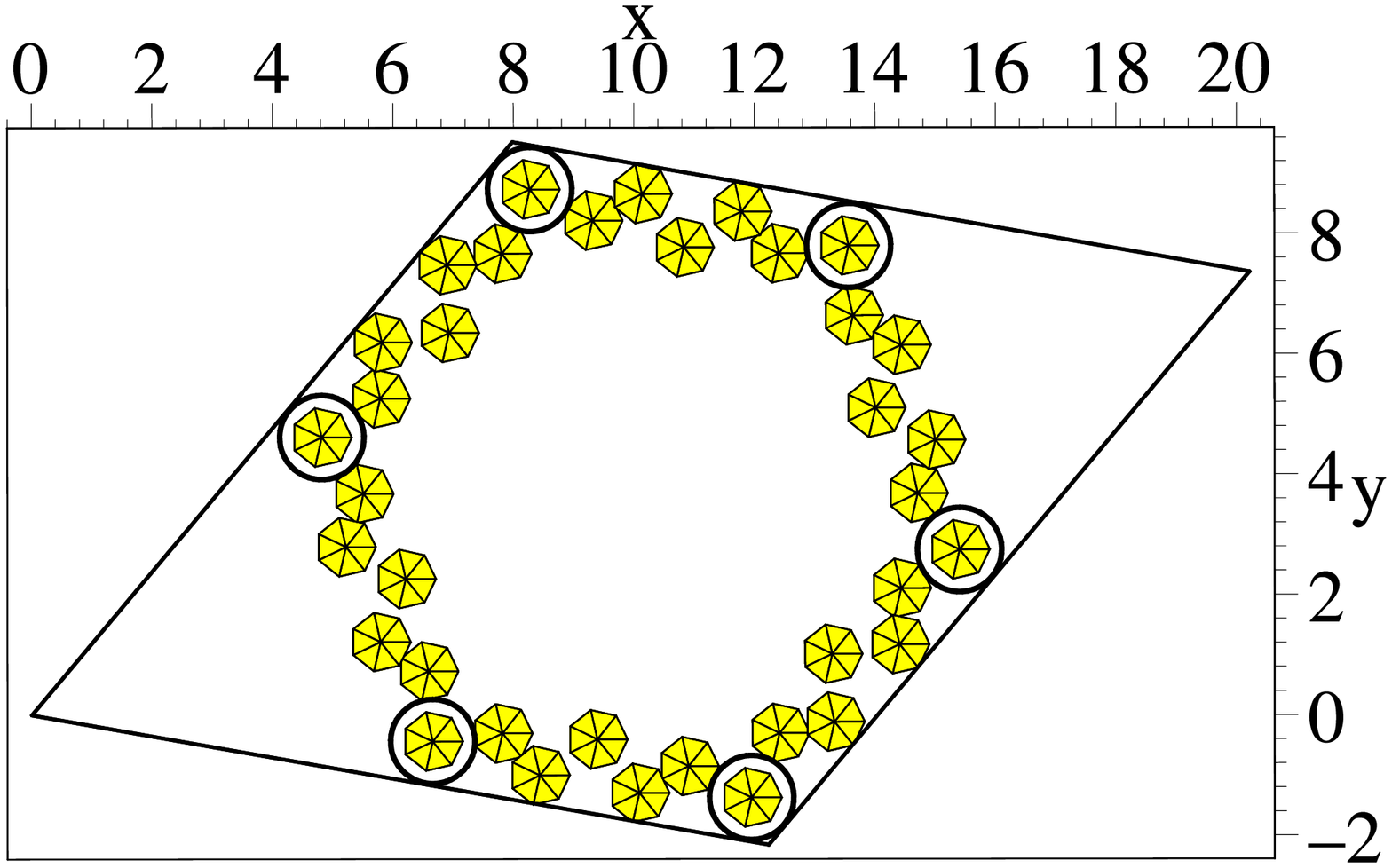}}

\parbox{\myw}{(c) Zigzag (6,0) CNT}
\parbox{\myw}{(d) Zigzag (6,0) BNT}
\subfigure{\includegraphics[angle=-90,width=\myw]{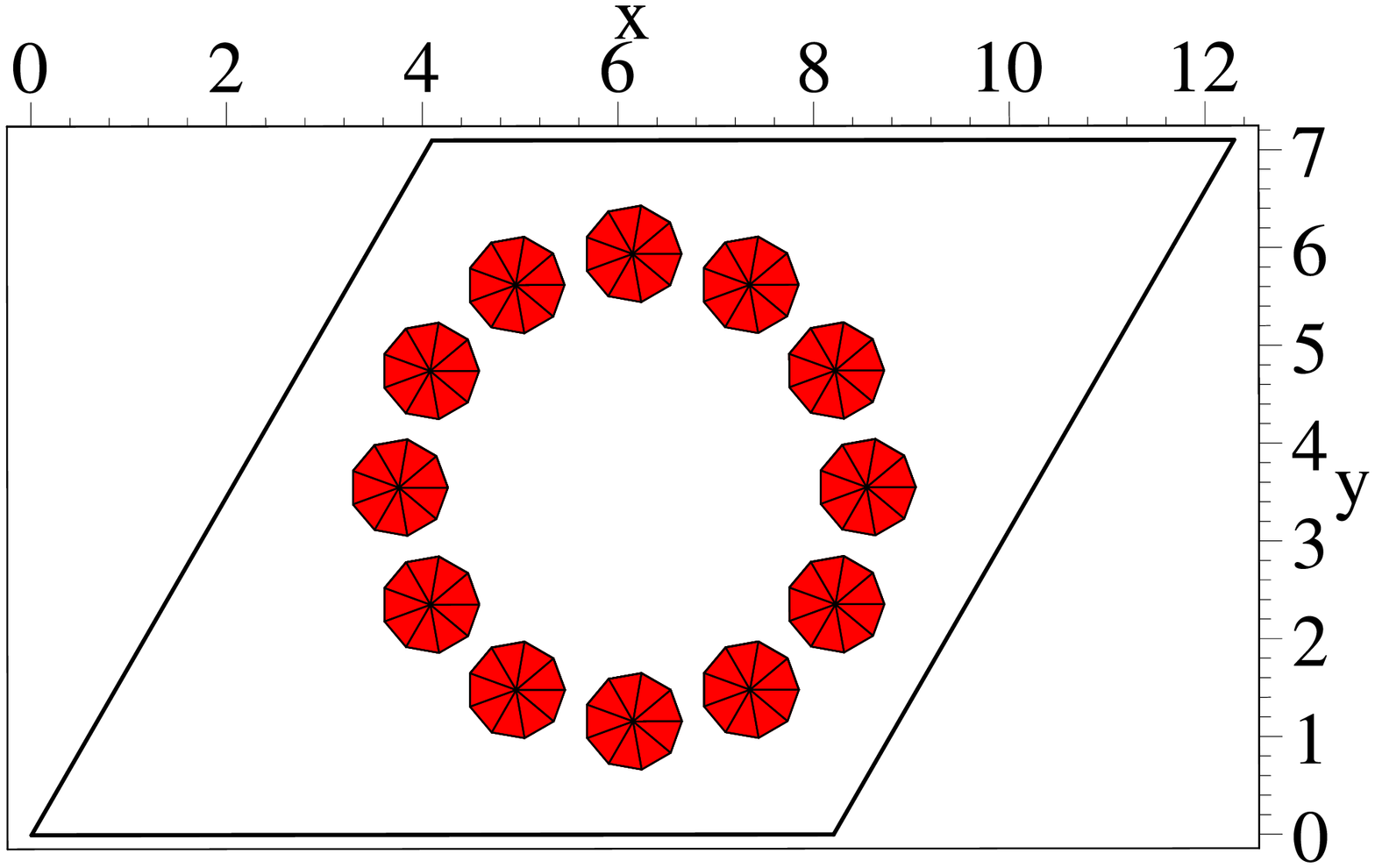}}
\subfigure{\includegraphics[angle=-90,width=\myw]{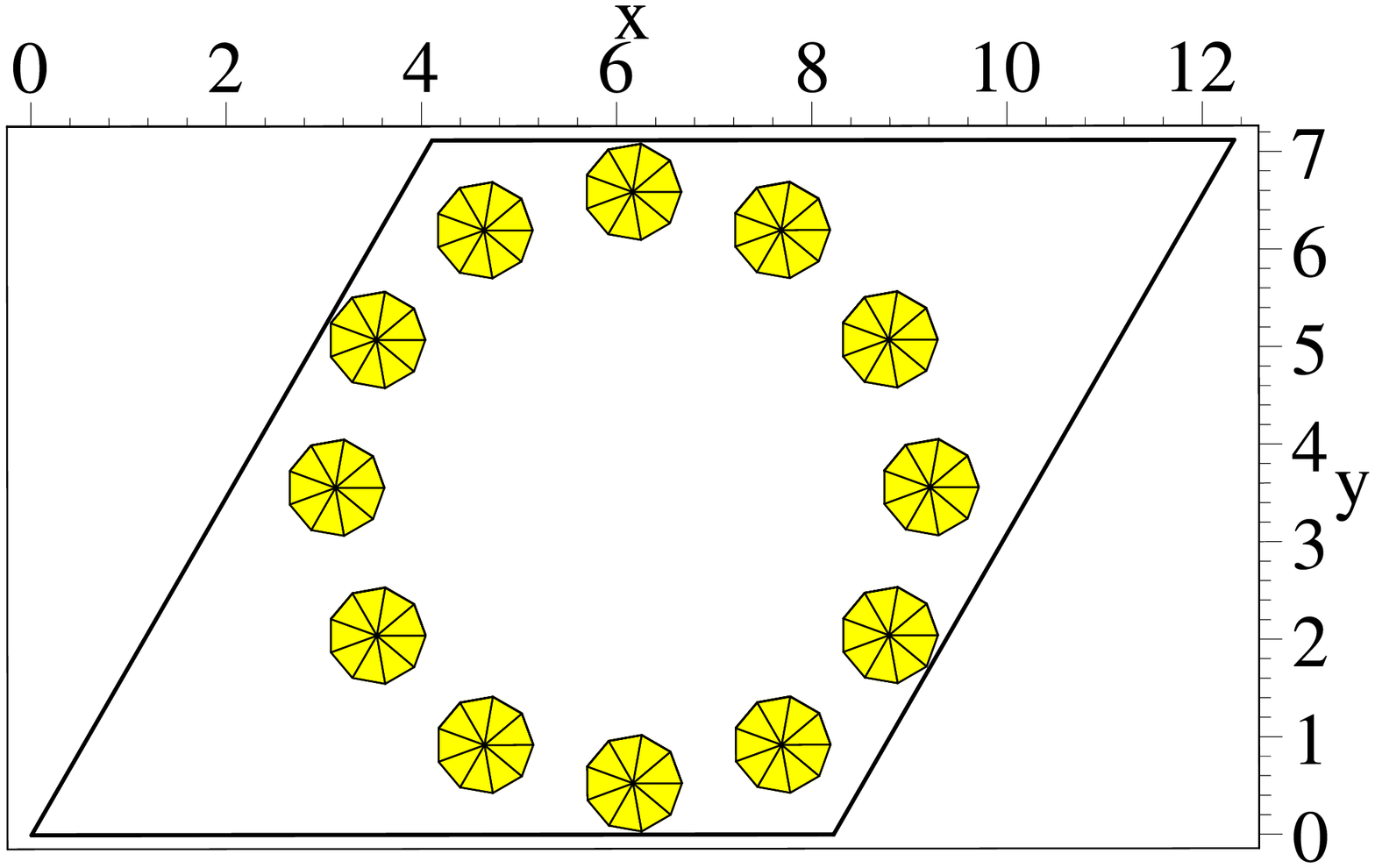}}
\caption{\label{fig:RefStr}
Top view of the (relaxed) reference structures placed around the center of 
their unit cells. Encircled atoms in (b) have additional covalent bonds to 
neighboring nanotubes on the superlattice. Units are \AA, light atoms are 
boron and dark atoms are carbon.}
\end{figure}

In the following, we will discuss our results for the following types of 
model junctions: Armchair 1 (A1) and Armchair 2 (A2)
approximate an armchair (6,6) junction, and Zigzag 1 (Z1)
approximates a zigzag (6,0) junction (Fig. \ref{fig:struc}). 
Besides (6,6) and (6,0) BNTs and CNTs have been relaxed, which are 
considered as suitable reference systems (Fig. \ref{fig:RefStr}).

A1 was prerelaxed with the Brillouin zone being sampled
by just the central $\Gamma$-Point, and finished with a 4x4x4
grid. 
A 4x4x4 grid was used to finish Z1, with smaller FFT-grids
and a smaller plane wave basis set to do the prerelaxation. 
Finally A2 was prerelaxed with the $\Gamma$-Point, only and finished 
with a 2x2x2 grid. 
The (6,0) CNT was relaxed with a 4x4x7 grid and the 
total energy was calculated on a 5x5x7 grid. The armchair 
(6,6) CNT was prerelaxed with 4x4x6 and finished with 5x5x9. 
The boron (6,0) zigzag nanotube was started with 3x3x5 and finished 
with 5x5x11 $k$-points. The relaxed armchair BNT structure was obtained 
from \cite{quandt:01}, where it was relaxed on a 7x7x7 grid. The cutoff 
energy for the expansion of the wave function in plane waves was 
286.6 eV for A2, 358.2 eV for A1, Z1 and the CNTs, and 321.4 eV for the BNTs.


The cohesive energy of each structure (Tab. \ref{tab:data}) was calculated 
by dividing the binding energy per unit cell by the number of atoms 
contained within that unit cell. In order to judge the stability of 
our model junctions, we decided to compare the cohesive energies of 
the relaxed structures to a reference cohesive energy of a phase-separated 
system of pure CNTs and pure BNTs of the same structure type, see
Table \ref{tab:data}. For the (6,6)-junction the reference binding
energy would be $E_{\mathrm{bind}}^{\mathrm{ref}}=24 \times 10.023
+ 36 \times 7.004 =492.696$ eV, and the corresponding reference cohesive 
energy would be $E_{\mathrm{coh}}^{\mathrm{ref}} =
E_{\mathrm{bind}}^{\mathrm{ref}}/60 = 8.212$ eV/atom. 
All cohesive energies as well as energetic differences with respect 
to the reference cohesive energies are given in table \ref{tab:data}.
 


\section{Results}
Our survey started with the relaxation of the reference structures, which are 
bundles (ropes) of pure CNTs and BNTs. They are displayed in 
Fig. \ref{fig:RefStr} (top view). While the surface of small radius CNTs is 
always smooth (Fig. \ref{fig:RefStr}a,c), BNTs \textit{can} exhibit a puckered 
surface \cite{boustani:99}, which shows up in Fig. \ref{fig:RefStr}b, but not 
in Fig. \ref{fig:RefStr}d. Due to higher B-B bond lengths, the average radius 
of a BNT is always bigger than the radius of a CNT of the same structure type. 
Another major difference between CNTs and BNTs is the potential of the latter 
to form covalent intertubular bonds \cite{quandt:01}. In contrast, the CNTs 
may only bind to each other via van der Waals types of interactions 
\cite{dresselhaus:SFCN}. Such covalent bonds are indeed present in the model 
(6,6) BNT, and the atoms that form these bonds are encircled in 
Fig. \ref{fig:RefStr}b. The coordination number of boron atoms in BNTs is 
usually six \cite{boustani:98} or even seven for atoms forming intertubular 
bonds, and it is always three in CNTs. The range of bond lengths, the 
parameters of the unit cells, and the cohesive energies of all fully relaxed 
structures can be found in table \ref{tab:data}.

\begin{figure*}
\setlength{\myw}{.3\linewidth}

\parbox{\myw}{(a) A1 - unrelaxed}
\parbox{\myw}{(b) A1 - relaxed}
\parbox{\myw}{(c) A1 - top view}
\subfigure{\includegraphics[angle=-90,width=5.5cm]{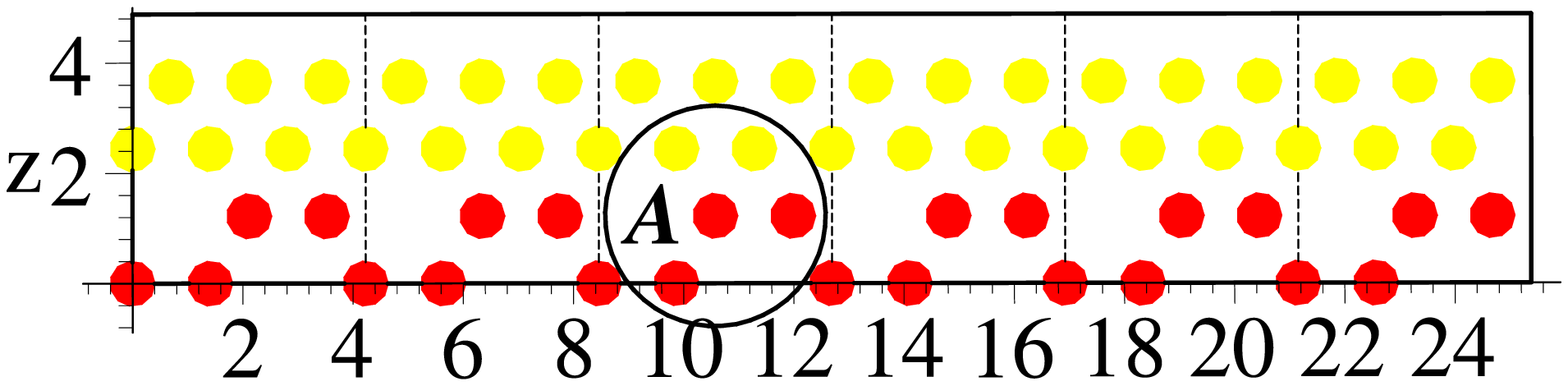}}
\subfigure{\includegraphics[angle=-90,width=5.5cm]{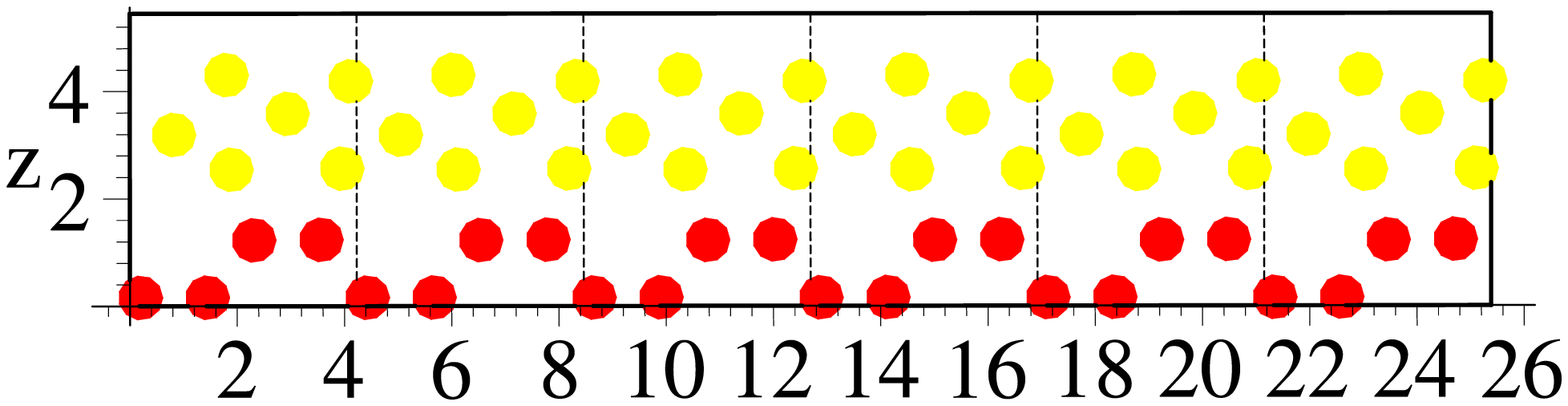}}
\subfigure{\includegraphics[angle=-90,width=4.5cm]{pics/A1-rlx-top.ps}}

\parbox{\myw}{(d) Z1 - unrelaxed}
\parbox{\myw}{(e) Z1 - relaxed}
\parbox{\myw}{(f) Z1 - top view}
\subfigure{\includegraphics[angle=-90,width=3.3cm]{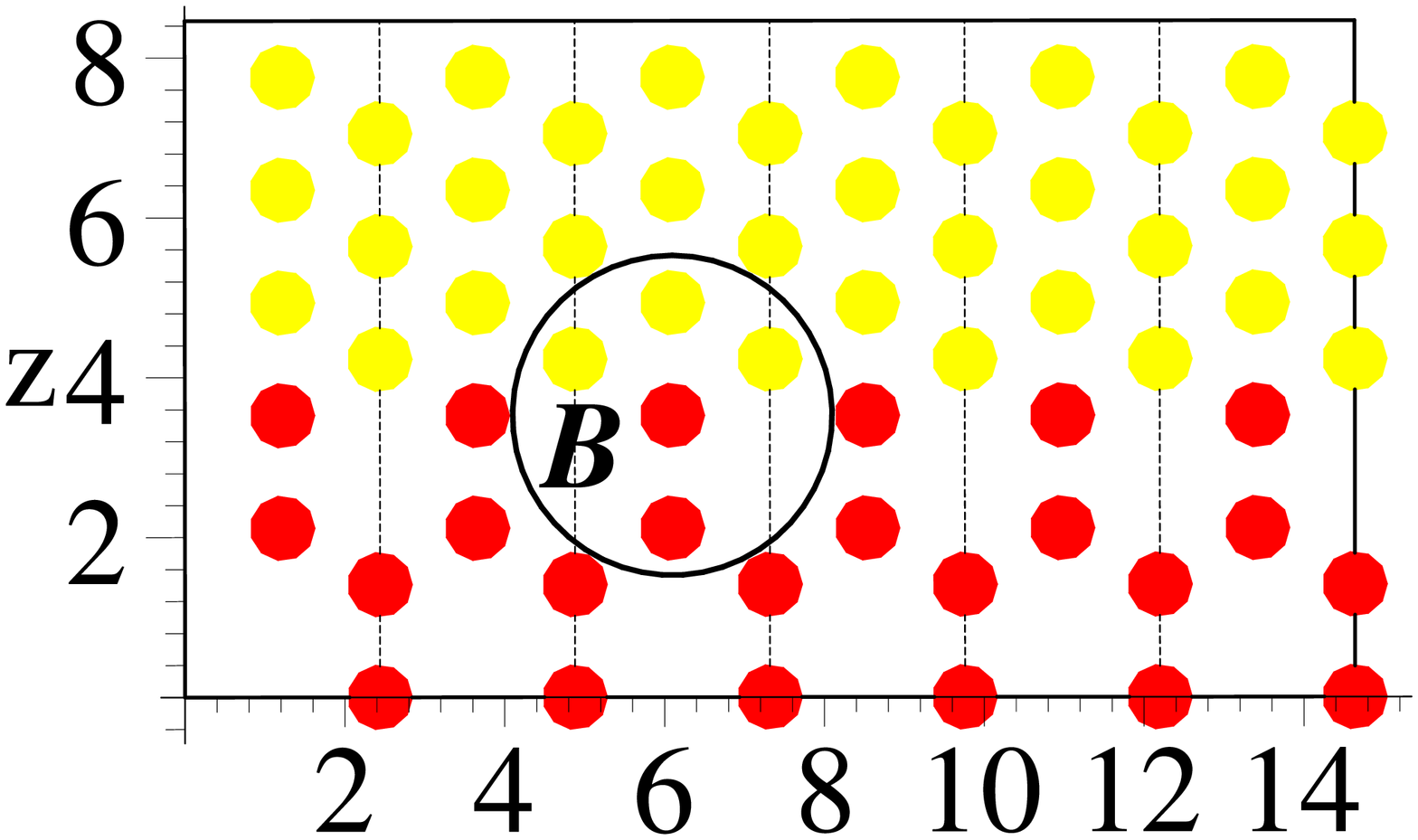}}
\hspace{2.2cm}
\subfigure{\includegraphics[angle=-90,width=3.3cm]{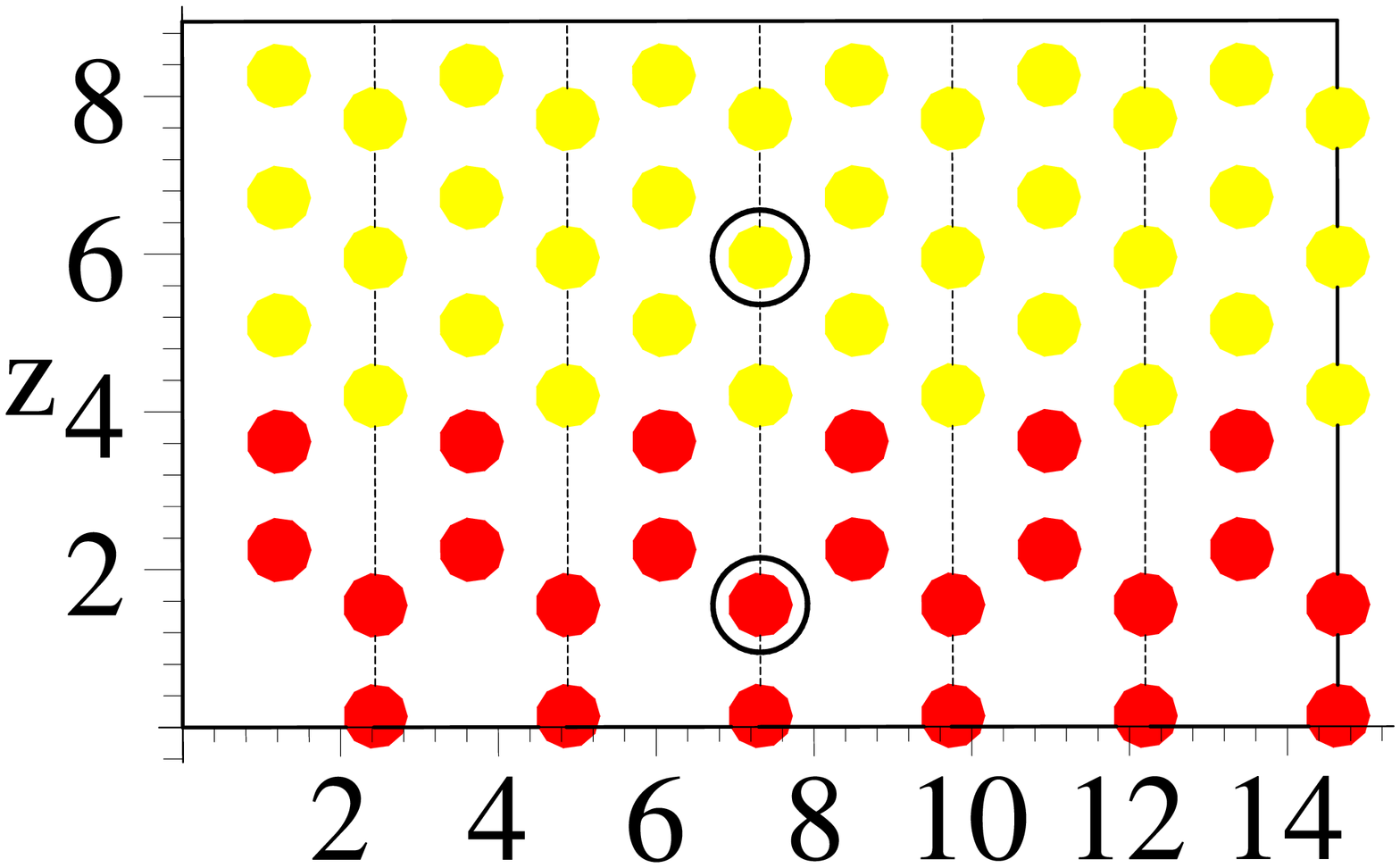}}
\hspace{2.2cm}
\subfigure{\includegraphics[angle=-90,width=4.5cm]{pics/Z1-rlx-top.ps}}

\parbox{\myw}{(g) A2 - unrelaxed}
\parbox{\myw}{(h) A2 - relaxed}
\parbox{\myw}{(i) A2 - top view}
\subfigure{\includegraphics[angle=-90,width=5.5cm]{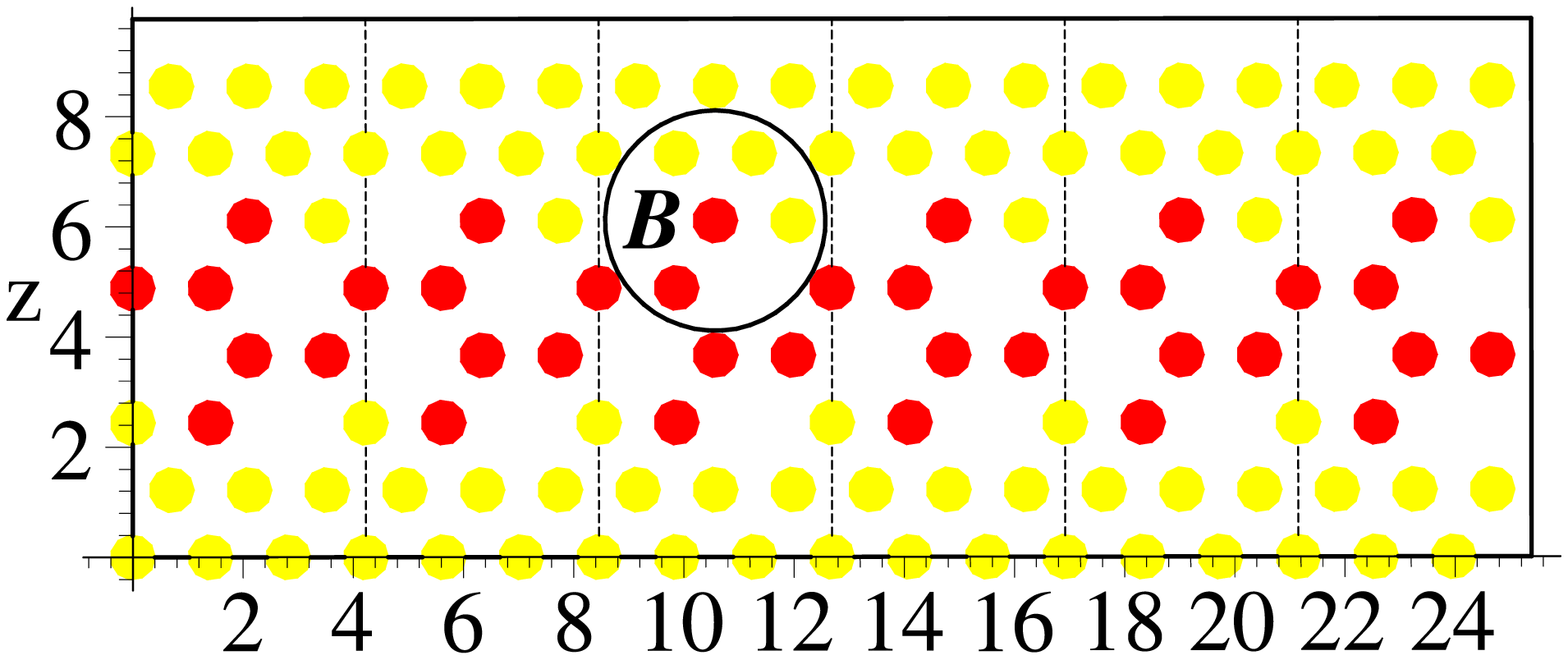}}
\subfigure{\includegraphics[angle=-90,width=5.5cm]{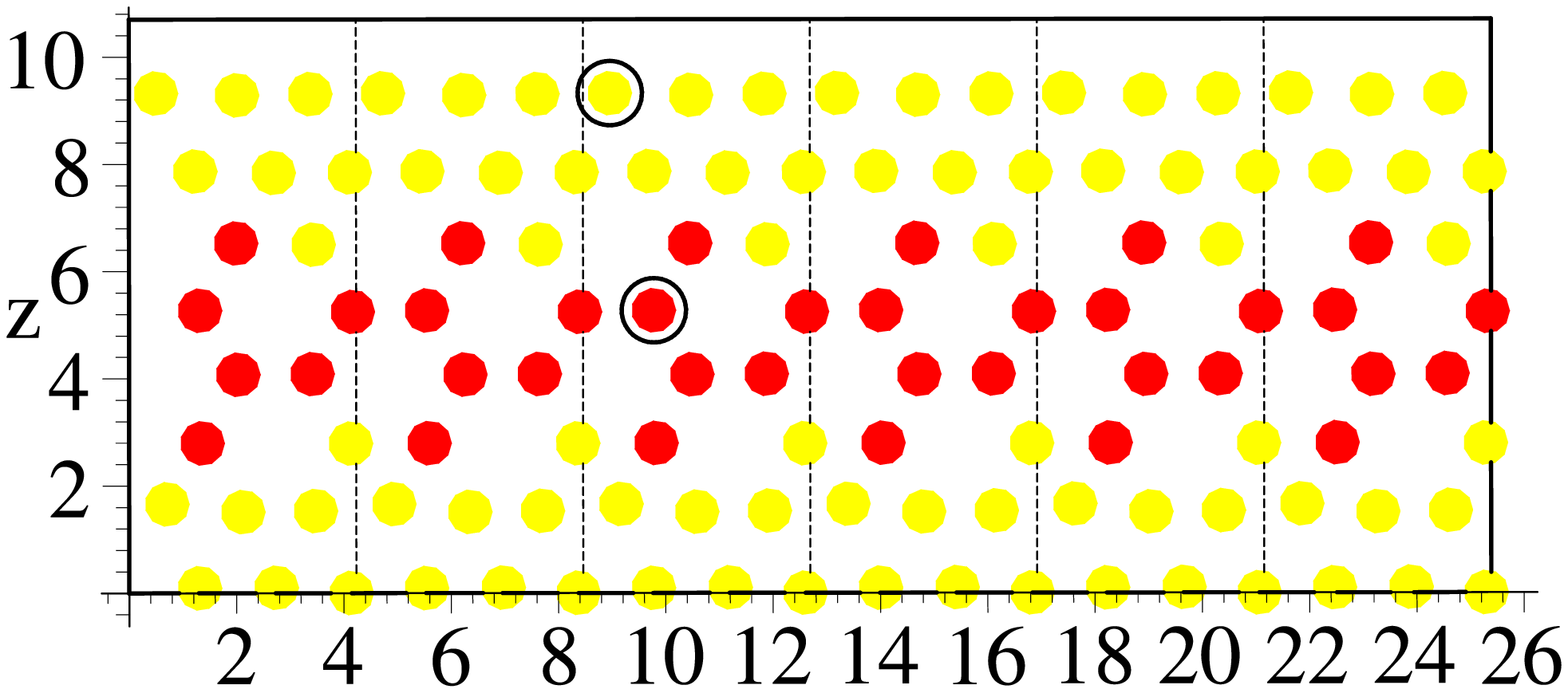}}
\subfigure{\includegraphics[angle=-90,width=4.5cm]{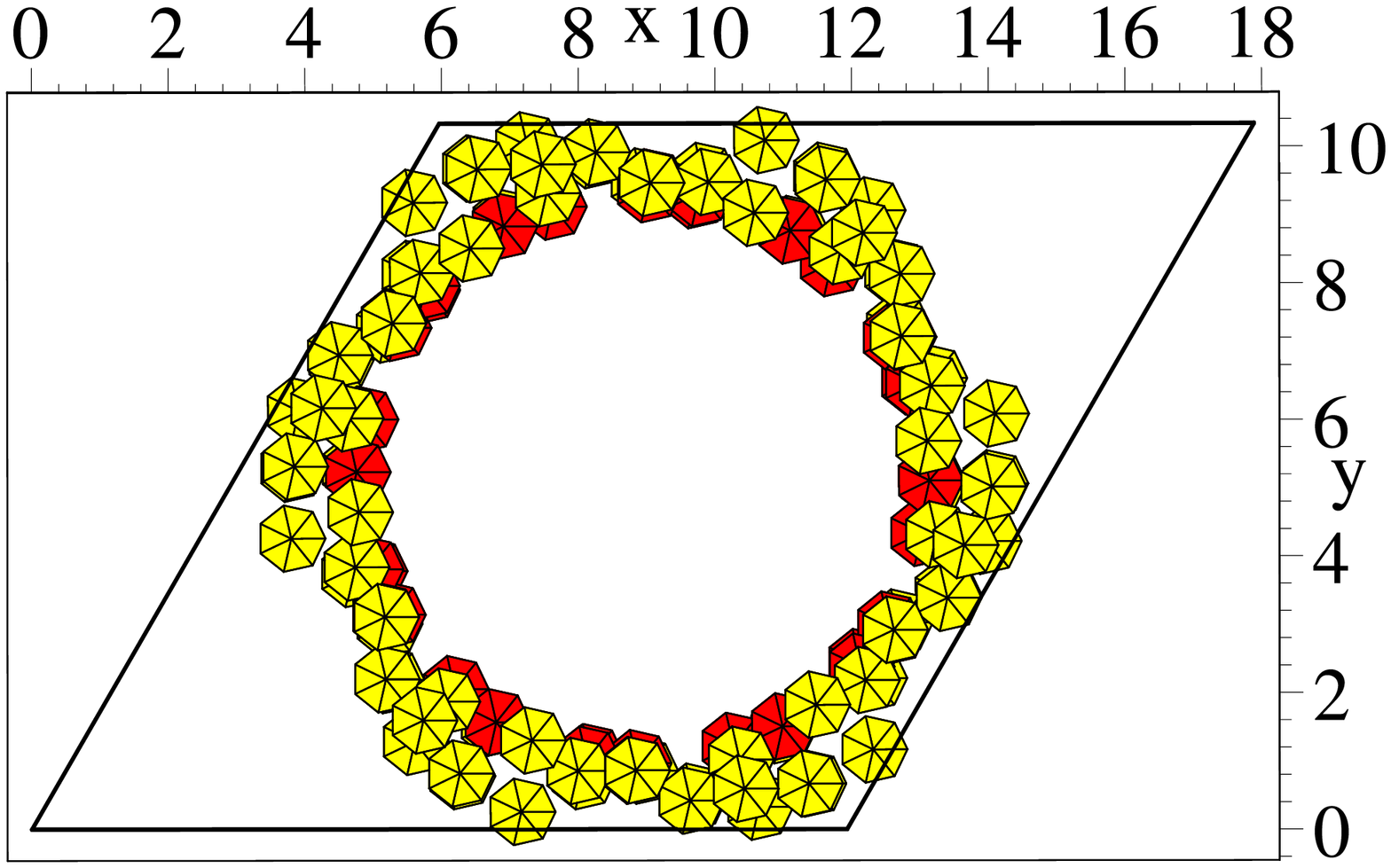}}
\caption{\label{fig:struc}
First column: starting geometries in planar view, encircled parts 
emphasize the structure elements $A$ and $B$. Dashed lines indicate 
six similar wedges partitioning the unit cells. Second column: relaxed 
structures in planar view. The local density of states of the encircled 
atoms in (e) and (h) is plotted in Fig. \ref{fig:dos}. Third column: top 
view of relaxed structures in their unit cells. Units are in \AA, and 
light atoms are boron, and dark atoms are carbon.}
\end{figure*}

The relaxed approximants A1, A2, and Z1 as well as the initial configurations 
are shown in planar view in Fig. \ref{fig:struc}. The left column shows the 
sheet, from which we constructed the starting configurations. In the middle 
the optimized junctions were projected back onto a similar sheet. The right 
column shows the junctions and their unit cells in top view.
All of these structures have a $C_6$-Symmetry, e.g. they may be dissected 
into six identical $60^{\circ}$-wedges (indicated by dashed lines in the 
planar views)\footnote{This $C_6$-Symmetry is generated by the six-fold 
symmetry of the hexagonal tubular superlattice as well as the structure 
types of the reference nanotubes, which are either (6,6) or (6,0)}. 

The cohesive energies of all approximants are below the cohesive energies of 
the corresponding phase-separated reference systems (see Tab. \ref{tab:data}). 
But the systems of isolated BNTs and CNTs are only sightly more stable than 
the junctions presented in this study.
Each junction on the superlattice will bind to neighboring tubes via 
covalent boron atoms. There are no covalent intertubular bonds involving 
carbon. Instead, the carbon part of the junctions almost rigidly stays in 
its initial geometry, and generally leads to smooth surfaces for the carbon 
part of the junctions. The boron on the other hand turns out to be more 
flexible, always forming puckered surfaces. 

We studied the properties of structure element $A$ by setting 
up the approximants A1. The cohesive energy of A1 lies 0.13 eV/atom 
below its cohesive reference energy, making this system slightly less
stable than the phase separated reference system.
After relaxation the hexagonal symmetry of the boron part has
disappeared, and the junction was transformed into a new structure
with 4-fold intratube coordination for boron, and 3-fold
coordination for carbon (see Fig \ref{fig:struc}b). It seems 
that carbon tends to retain the planar $sp^2$-environment and 
forces the boron atoms into an unusual quasiplanar arrangement 
with rather low coordination numbers. 

To overcome the somewhat constrained arrangements of A1, we set 
up another approximant (not shown) chosen to be twice as long in 
the $z$-direction, and comprising twice as many atoms, but still 
containing structure element $A$, only. The relaxed structure had 
the same structural problems as A1 and was almost isoenergetic to 
A1. This gives a consistent picture and certainly shows that the 
structural problems encountered for these types of armchair 
approximants are no artefacts of small unit cells. Thus the 
starting geometries involving the structure element $A$ are obviously 
far from optimal.

In a next step we studied the structure element $B$ by surveying approximant Z1. 
Its cohesive energy lies just 0.07 eV/atom below the reference energy.
The planar view in Fig. \ref{fig:struc}e looks almost identical to its starting 
configuration and the basic lattice structures in both parts of the junction 
are maintained, allowing for high coordination numbers of the constituent atoms. 
It seems as if this type of heterojunction could be formed by sticking together 
two perfect zigzag tubes. 
It therefore looks as if nanotubular junctions formed with structure element $B$ are generally favorable.

Therefore we tried to incorporate $B$ in junctions of armchair type as well. 
To this end, one has to modify the construction scheme and allow for a 
mixing of the species at the interface. Such a scenario has been realized 
in approximant A2 (Fig. \ref{fig:struc}, third line). 
The unit cell has been shifted by $c/4$. Now the two transition regions in every 
unit cell are clearly visible. The relaxed junctions shows properties very 
similar to Z1. Its cohesive energy is 0.07 eV/atom below the reference energy. 
Again the planar view in Fig. \ref{fig:struc}h looks almost identical to its 
initial configuration, indicating a starting point on the energy hypersurface 
very close to the sought mini\-mum. 
Though very different in geometry Z1 and A2 exhibit very similar properties: 
the same stability, the same mean coordination number and the same mean 
number of intertubular bonds. 

\begin{figure}
\setlength{\myw}{.4\linewidth}

\centering
\subfigure{\includegraphics[angle=-90,width=\myw]{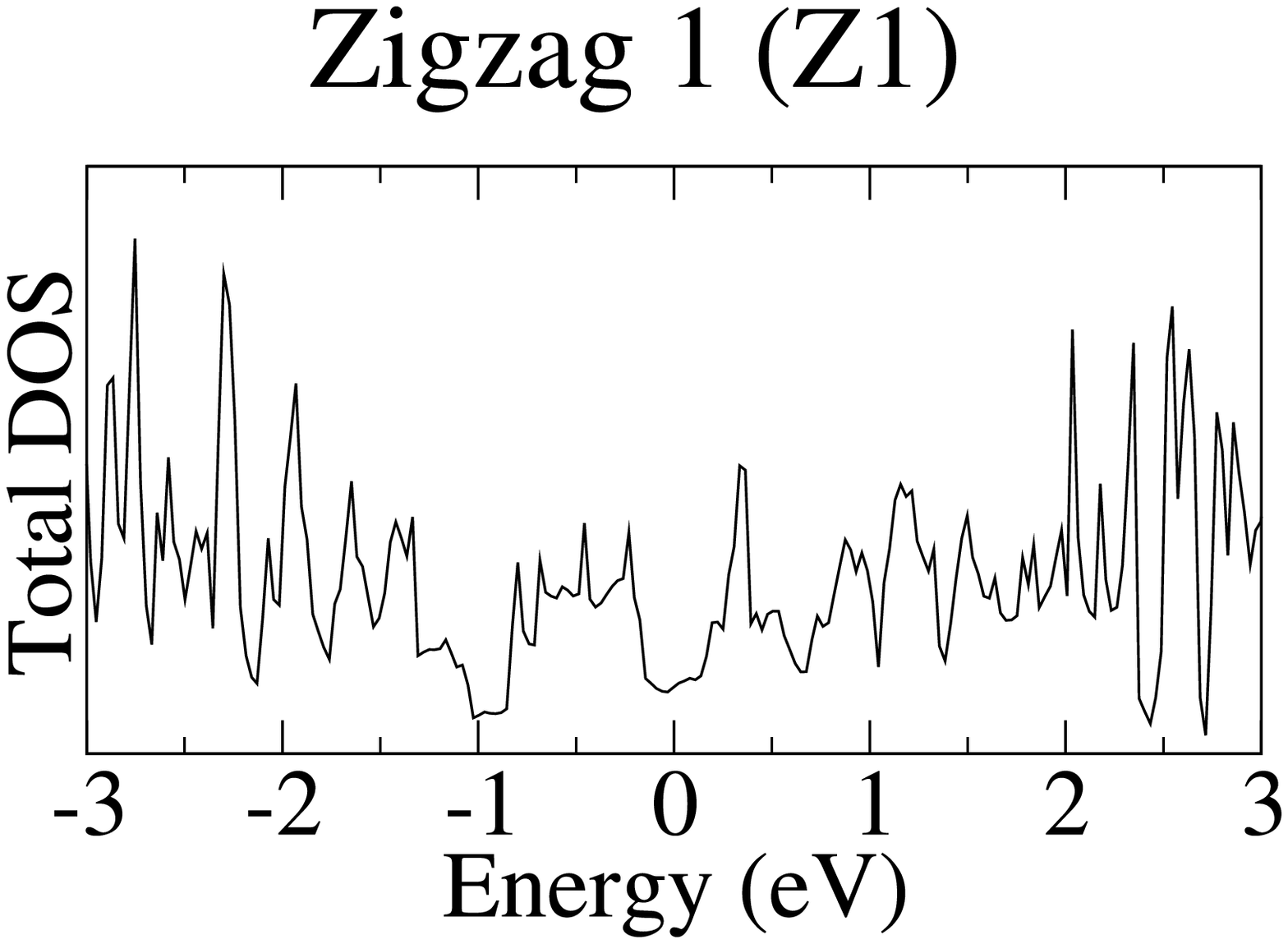}}
\hspace{0.05\linewidth}
\subfigure{\includegraphics[angle=-90,width=\myw]{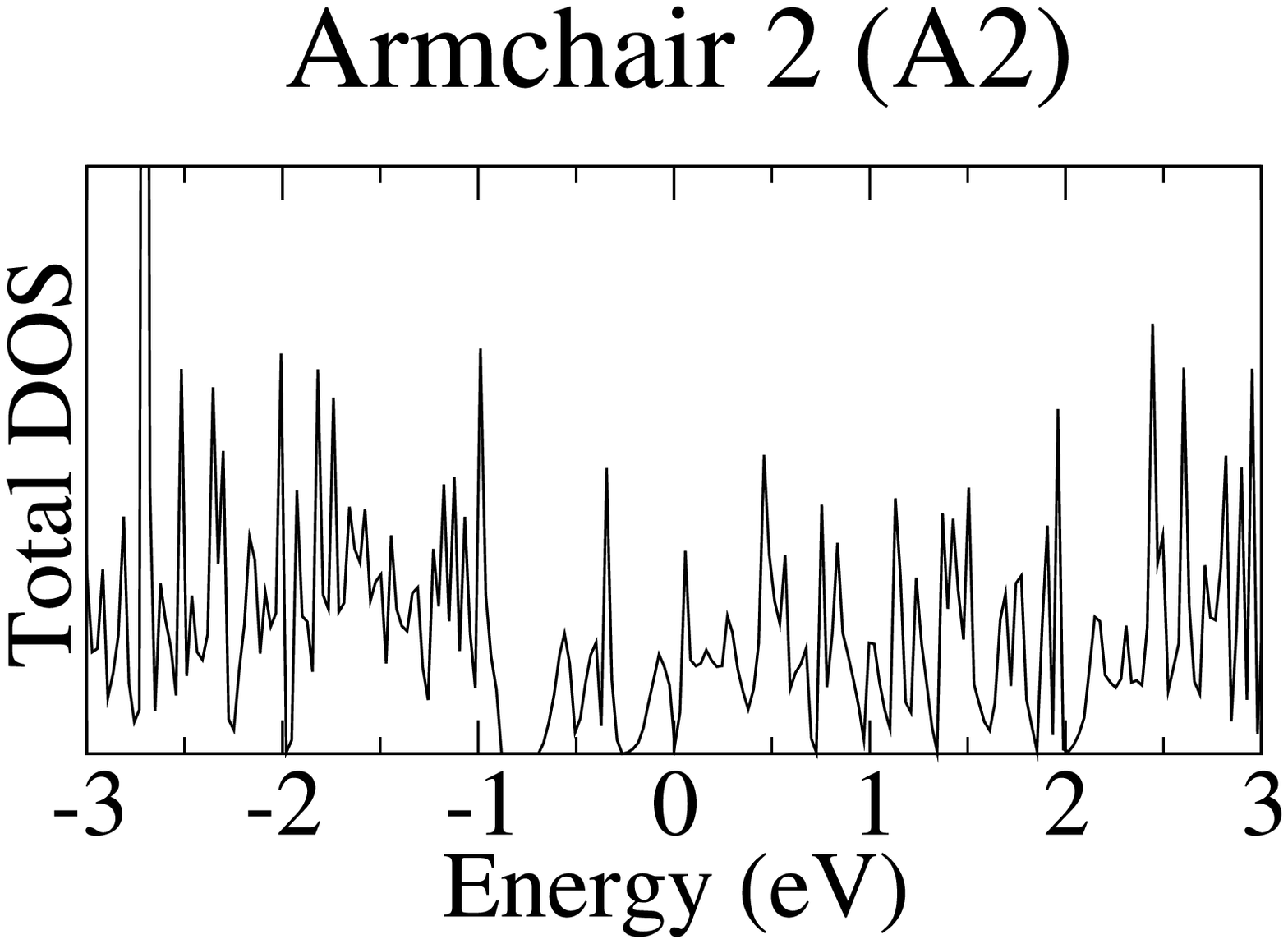}}

\subfigure{\includegraphics[angle=-90,width=\myw]{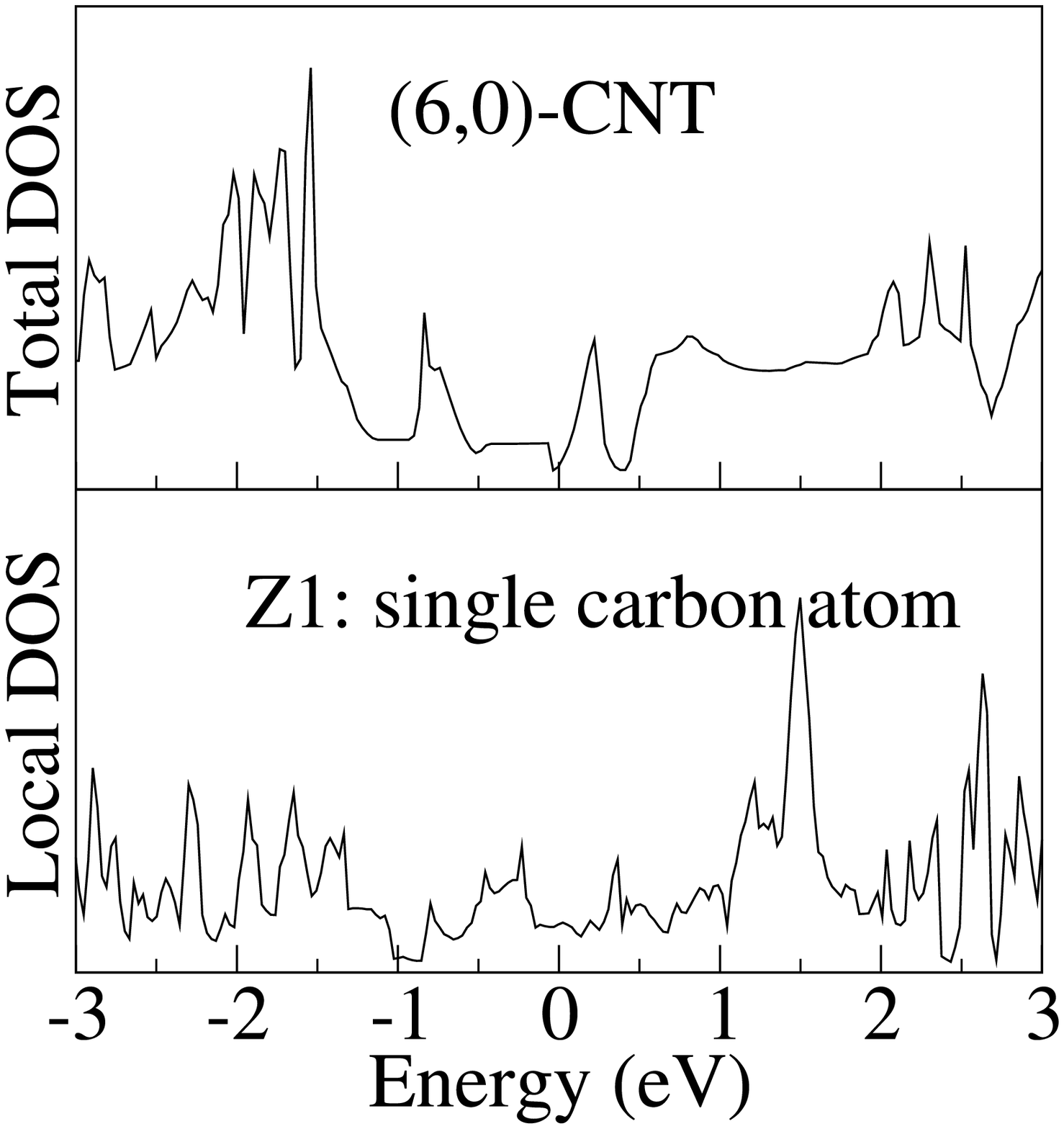}}
\hspace{0.05\linewidth}
\subfigure{\includegraphics[angle=-90,width=\myw]{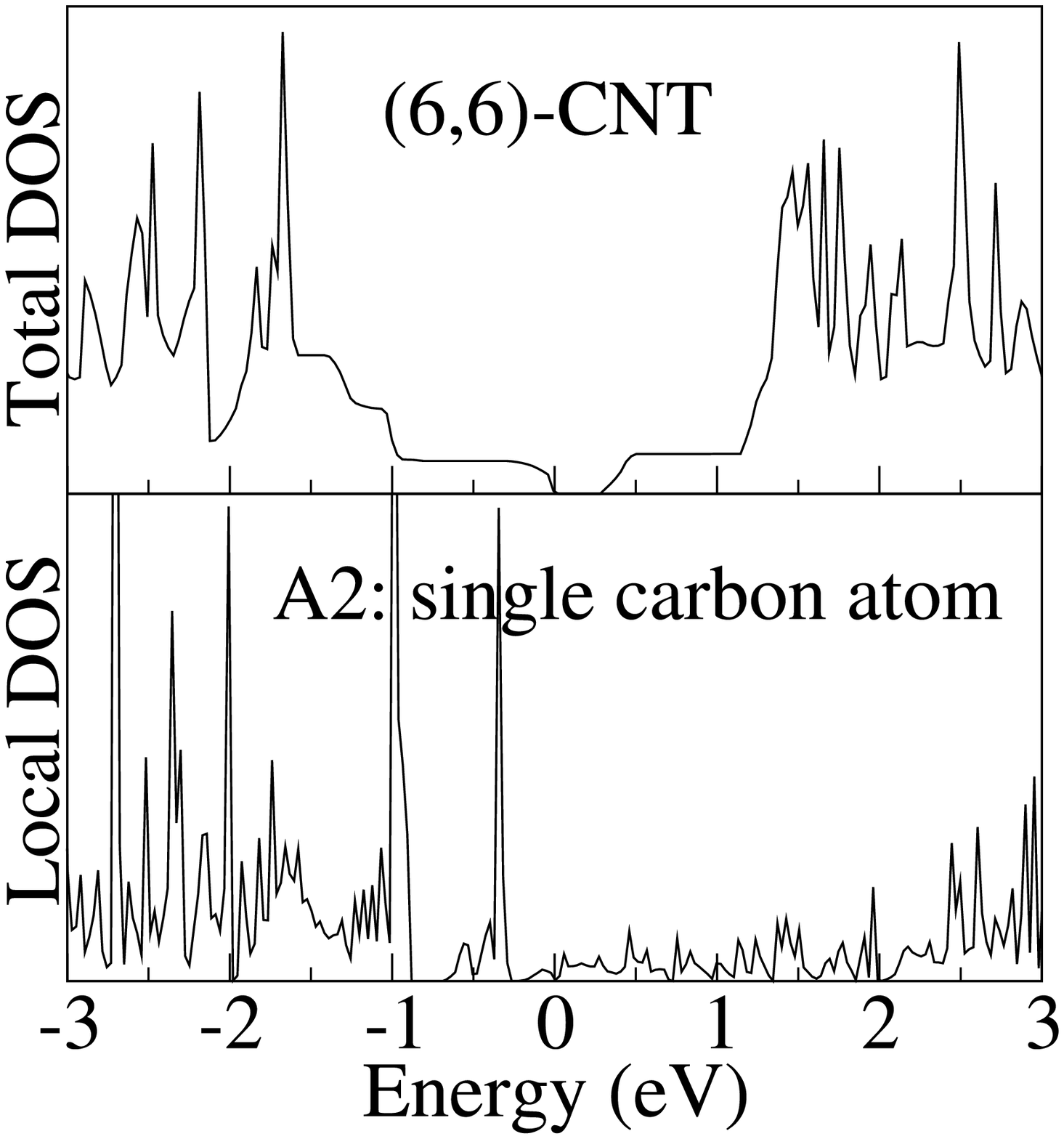}}

\subfigure{\includegraphics[angle=-90,width=\myw]{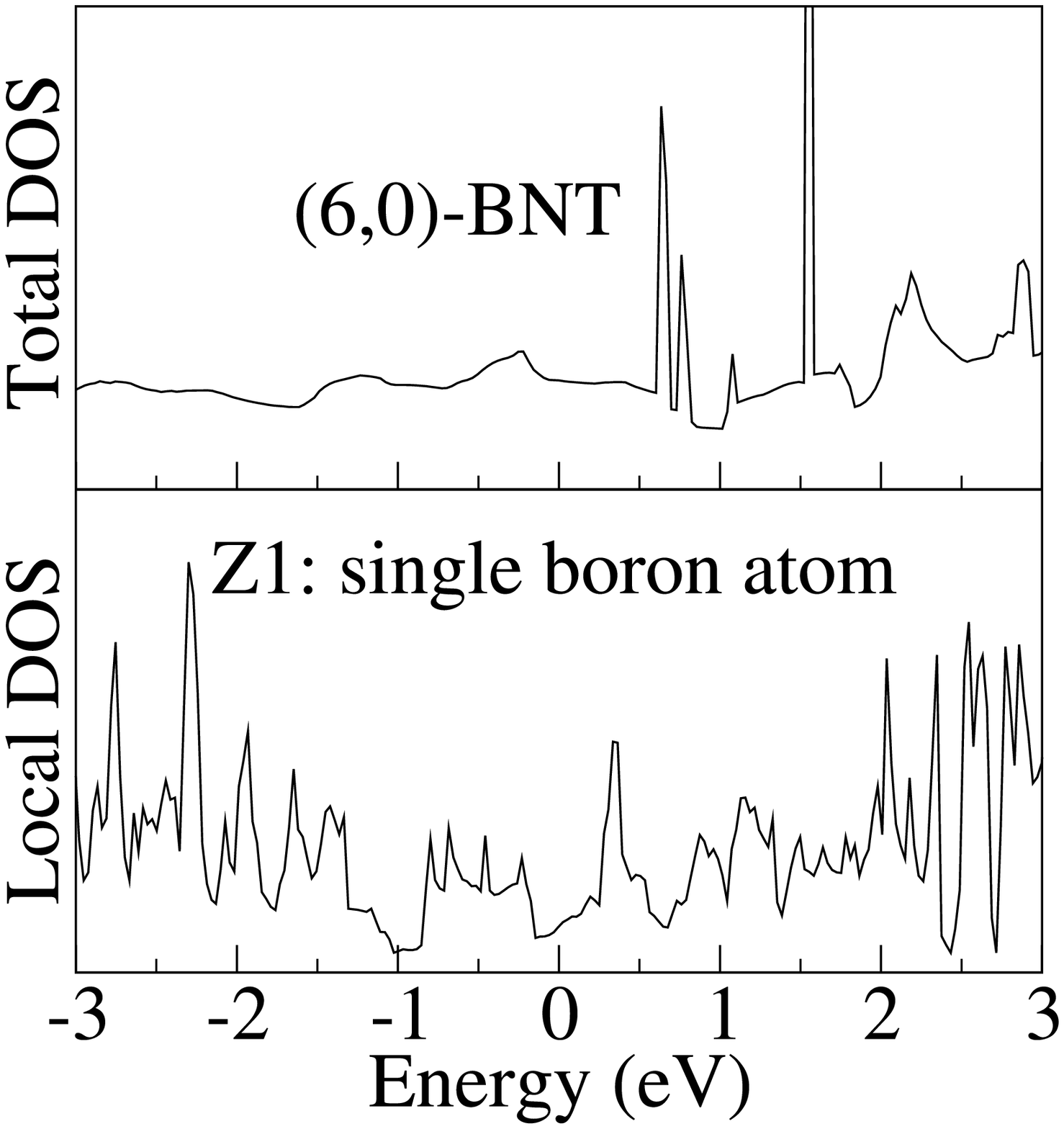}}
\hspace{0.05\linewidth}  
\subfigure{\includegraphics[angle=-90,width=\myw]{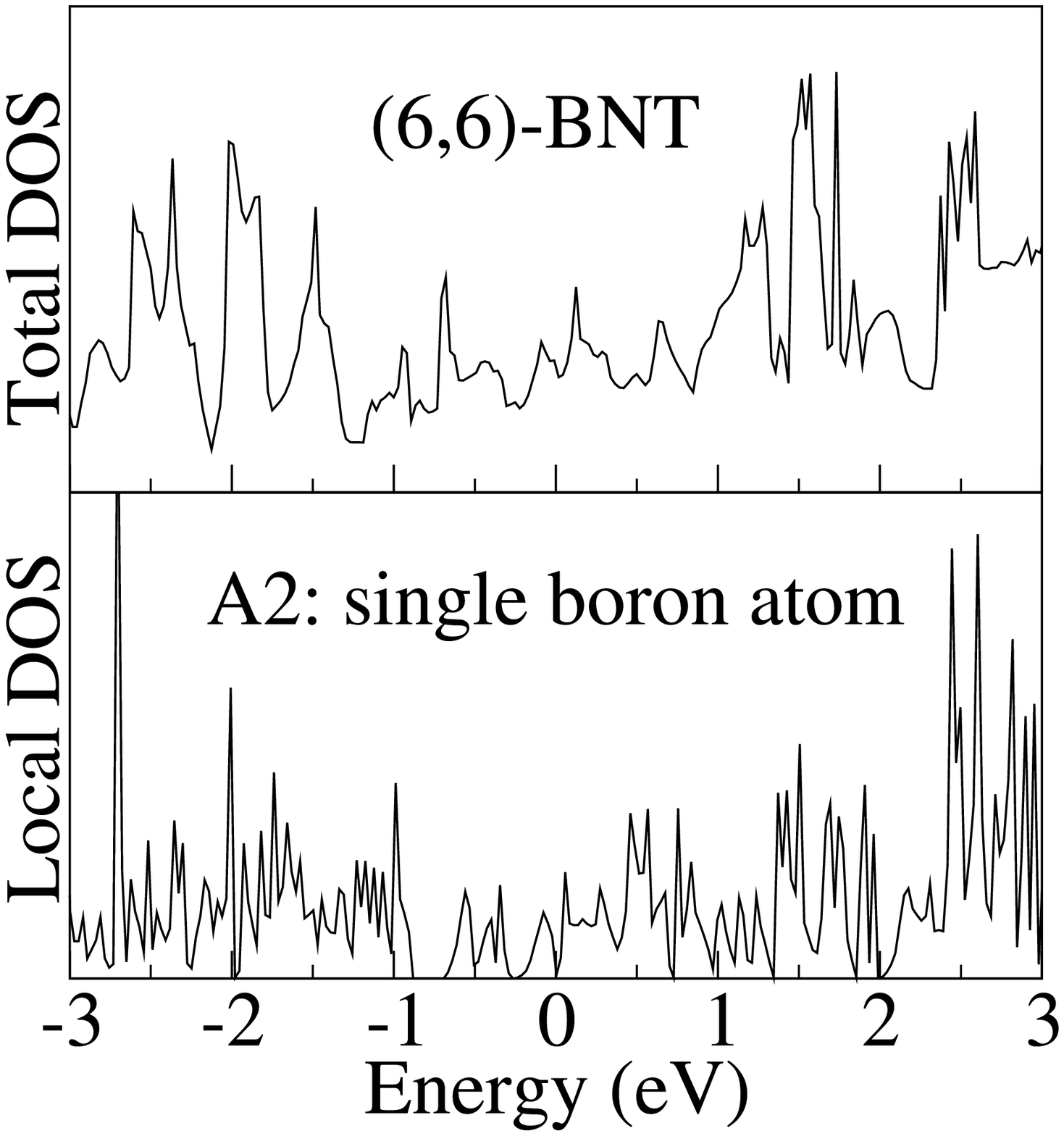}}  
\caption{\label{fig:dos}
Total and local density of states (DOS) in arbitrary 
units of Z1 and A2 (first and second column, respectively). In the split-panels we compare the total DOS of the reference structures (bulk) (see Fig. \ref{fig:RefStr}) to local DOS of atoms within the junctions. These atoms are encircled in Fig. \ref{fig:struc}e and \ref{fig:struc}h. $E=0$ equals the Fermi energy $E_F$.}
\end{figure}

Finally we would like to discuss the basic electronic properties of Z1 and A2.
The first line of Fig. \ref{fig:dos} shows the total electronic density of 
states (DOS) of the structures near the Fermi energy $E_F$. 
The double-plots allow for a comparison of the local DOS of a central 
C- or B-atom (these atoms are encircled in Fig. \ref{fig:struc}e and 
\ref{fig:struc}h)\footnote{Due to the $C_6$-Symmetry the local DOS of 
similar atoms within different wedges are (almost) identical, while 
non-equivalent atoms differ significantly.} to the DOS of the reference 
systems, which are considered as bulk. Both are very different. Thus the 
approximants obviously do not contain bulk states and the transition to 
the bulk proceeds over wider ranges, which cannot be simulated by the small 
approxi\-mants presented in this study.
The position of major van-Hove-peaks and band gaps in the local DOS is 
homogeneous throughout the junction (thus it does not differ in the local 
and total DOS). This could be an indication for a common $\pi$-electron 
system extending throughout the junction. Finally we also found that the 
total DOS looks more similar to the local DOS of the boron atom rather 
than to the local DOS of carbon. 

\section{Conclusions}
For the first time we explored the stability, geometry and basic chemical and 
electronic properties of sample carbon-boron heterojunctions. 

Starting from a general prescription to generate linear nanotubular 
junctions with arbitrary radius and chirality we showed that only two 
structure elements $A$ and $B$ can appear at the B-C transition of such 
a junction. The general atomic arrangements of $A$ and $B$ look rather 
similar, but in $A$ the central carbon atoms will bind to two boron and 
two carbon atoms, whereas for $B$ the central carbon atom will bind to 
three boron and one carbon atom, only. 

We then studied the properties of $A$ and $B$ independently by examining 
approximants for a (6,6) armchair and a (6,0) zigzag junction. From a 
structural point of view, the carbon part of all relaxed junctions looks 
rather similar to what is known for stand-alone carbon nanotubes - exhibiting 
a smooth surface and preferring carbon in a 3-fold coordinated environment 
\cite{dresselhaus:SFCN}. The boron part of the junctions is always puckered 
and forms intertubular bonds, as known from stand-alone boron nanotubes 
\cite{boustani:97,boustani:98}, but the ideal hexagonal arrangement 
in the boron part of the junctions is only preserved by junctions that 
feature the structure element $B$. Such systems have similar properties; 
they display the same average coordination numbers, the same types of 
intertubular bonds, and they are only 0.07 eV/atom less stable than a 
phase separated system of carbon and boron nanotubes. Junctions that 
feature element $A$ have structural problems and their formation energies 
per atom are approximately twice as high as for junctions formed with 
element $B$.

The central carbon atom of structure element $B$ is 4-fold coordinated. 
Besides the known tetragonal $sp^3$ arrangements, we found new planar 
and quasiplanar atomic arrangements induced by the electron deficiency of boron \cite{pauling_1960_ncb}.
Such forms have not been reported in solid system before and are in partial agreement with results found by Exner \textit{et al.} \cite{exner:00}, who studied small planar BC-clusters.

The analysis of the total and local electronic DOS 
indicates that all junctions consist of interface states, only. 
A transition towards the bulk will proceed over wider ranges than 
those that were studied here. Furthermore there are indications for 
a common $\pi$-electron system extending throughout the whole junction.

After relaxation all approximants were still tubular, had distinct boron and carbon parts and showed a clear B-C interface. This itself is a strong indication for the compatibility of these two nanotubular materials.
In detail, it seems that structure element $B$ is the key to  
generate optimal zigzag and armchair junctions, and those results also suggest
that \textit{any} stable linear and chiral nanotubular B-C junction may be 
formed by building structure elements $B$ within the transition region.
The relative instability of 0.07 eV/atom for the apparently optimal model 
junctions is surprisingly small, considering that at the interface both 
the boron and the carbon lattices are significantly disturbed, and second, 
that our approximants are really small. This energy barrier can be overcome 
thermodynamically, and it is very likely to decrease further for bigger systems. The influence of lattice perturbations is apparent in A1, where only small dislocations of the boron atoms give rise to a doubling of the relative instability. We thus conjecture that heterogeneous junctions between boron and carbon nanotubes should be possible. 
Our study also clarifies the \textit{local} chemistry at the interface of B-C
junctions, which could be shown to be independent of the periodic boundary 
conditions.

Finally, we hope that further studies of suitable model systems similar 
to the structures presented in this study will shed some light on the 
formation and stability of other nanotubular heterojunctions and networks.
According to our studies nanotubular carbon and boron materials seem to be compatible, 
and given the large number of structurally related materials (see introduction), 
our results also suggest the possibility of a large variety of novel 
heterogeneous nanotubular materials forming extended network systems. By 
taking advantage of the fortunate structural and electronic properties 
of heterogeneous nanotubular systems, one might even hope for the development 
of novel compound nano\-tubular devices in the future.

\section{Acknowledgments}
The authors thank Dr.\ Sylvio Kosse (Greifswald)
for technical support during our extensive use of the 'snowwhite' computer cluster, and Dr.\ Ihsan Boustani
(Wuppertal) and Prof.\ Klaus Fesser (Greifswald) for various
helpful discussions.

\bibliography{article}

\end{document}